\documentclass[prl,twocolumn,groupedaddress]{revtex4}
\usepackage{ae,aecompl}

\usepackage{graphicx}
\usepackage{amssymb}
\usepackage{amssymb}

\begin{document}

\title{Experimental observation of oscillating and interacting matter wave dark solitons}

\author{A. Weller}
\author{J.P. Ronzheimer}
\author{C. Gross}
\author{J. Esteve}
\affiliation{Kirchhoff Institute for Physics, University of
Heidelberg, INF 227, 69120 Heidelberg, Germany}

\author{D.J. Frantzeskakis}
\affiliation{Department of Physics, University of Athens, Panepistimiopolis, Zografos, Athens 157 84, Greece}
\author{G. Theocharis}
\author{P.G. Kevrekidis}

\affiliation{Department of Mathematics and Statistics, University of
Massachusetts, Amherst MA 01003-4515, USA}
\author{M.K. Oberthaler}

\affiliation{Kirchhoff Institute for Physics, University of
Heidelberg, INF 227, 69120 Heidelberg, Germany}
\email[]{dark_solitons@kip.uni-heidelberg.de}

\date{\today}

\begin{abstract}
We report on the generation, subsequent oscillation and interaction
of a pair of matter wave dark solitons. These are created by
releasing a Bose-Einstein condensate from a double well potential
into a harmonic trap in the crossover regime between one dimension
(1D) and three dimensions (3D). Multiple oscillations and collisions
of the solitons are observed, in quantitative agreement with
simulations of the Gross-Pitaevskii equation. An effective particle picture is developed and confirms that the deviation of the observed oscillation frequencies from the asymptotic prediction $\nu_z/\sqrt{2}$, where $\nu_z$ is the longitudinal trapping frequency, results from the dimensionality of the system and the soliton interactions.
\end{abstract}


\maketitle

Solitons are one of the most prominent features of nonlinear dynamics emerging in diverse fields extending from hydrodynamics to solid state physics and from
nonlinear optics to biophysics. Dark solitons are the fundamental excitations of the defocusing nonlinear Schr\"{o}dinger equation~\cite{ZS73}, and have the form of a localized ``dip'' on a background wave, accompanied by a phase jump~\cite{KLD98}. These localized waveforms have been demonstrated experimentally in different contexts, including liquids~\cite{Den90}, discrete mechanical systems~\cite{Den92}, thin magnetic films~\cite{Che93,Kal00}, optical media~\cite{emplit87,krokel88,weiner88,andersen90,swartzlander91}, and, more recently,  Bose-Einstein condensates (BECs)~\cite{Bur99,Den00,And01,dutton,engels,Jo07,Bec08}.
The possibility of creating pairs of dark solitons~\cite{krokel88,andersen90} has stimulated considerable interest in the repulsive~\cite{blow85} collisional interactions between them~\cite{Fou96,bang06,Bongs01}. The fundamental features of soliton collisions have a universal character and thus, e.g., optical solitons interact essentially the same way as matter-wave solitons.

In this letter we report on the systematic generation of a pair of matter wave dark solitons which is subsequently oscillating and colliding in a harmonic trap. Our experiment is performed in the crossover regime between 1D and 3D~\cite{Men02}, where dark solitons exist and are robust~\cite{Mur99}. This allows us to monitor, to our knowledge for the first time in any field, multiple oscillations and collisions of
dark solitons, permitting the precise measurement of their
oscillation frequency and their mutual repulsive interactions. Previous experiments have been performed in a genuine 3D regime where dark solitons are unstable due to the so-called snaking instability and eventually decay into vortex rings~\cite{Mur99,And01}. In these experiments solely their translation in the trap has been shown~\cite{Bur99,Den00,And01}. Only very recently dark solitons have been
reported to undergo a single oscillation period in a harmonic trap~\cite{Bec08}.

\begin{figure}[t]
    \centering
    \includegraphics[width=0.4\textwidth,keepaspectratio]{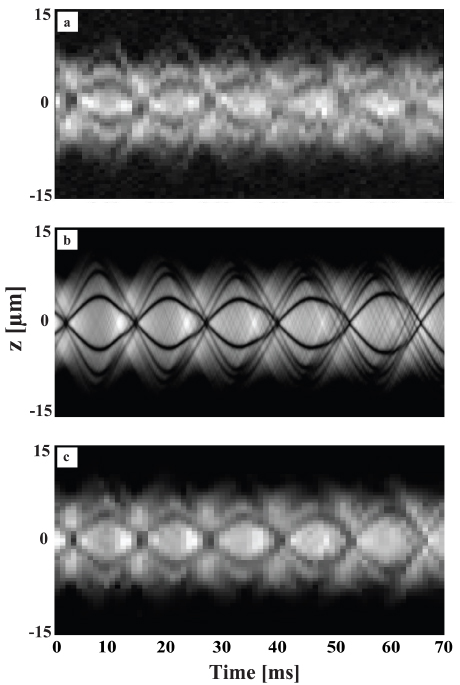}
    \caption{\label{fig1}
  Observation of the time evolution of dark solitons in a harmonic trap. a) Experimental observation of the dynamics of
    the longitudinal atomic density for the case of $N=1700$ atoms and trapping frequencies $(\nu_z,\nu_{\perp})$=(53 Hz,890 Hz) after a short time of flight. The shown images are averaged over 10 realizations of the experiment.
    b) Result of the numerical integration of the 3D GPE taking into account the full preparation process. c)
    The theoretical prediction taking into account the finite
    spatial as well as temporal resolution of the experiment.}
\end{figure}

Different methods have been explored to create dark solitons in Bose-Einstein
condensates~\cite{Bur99,Den00,And01,dutton,engels,Jo07,Bec08}.
In our experiment, the solitons are generated by merging two coherent condensates initially prepared in a double well potential. The observed evolution in the
trap, after the preparation process, is shown in Fig.~\ref{fig1}a revealing that an even number of solitons is generated. This formation process of the dark solitons can be regarded as a consequence of matter wave interference of the two condensates~\cite{RC97,Sco98,Lee07,SHO08}. Our procedure is very similar to the recently reported generation of vortices out of a triple well potential~\cite{Sch07}.

Since the two dominant solitons are created with a distance of a few
healing lengths, the repulsive interaction between them leads to a
significant modification of the oscillation frequency. The measured
frequencies deviate up to 16\% from the single soliton asymptotic
Thomas-Fermi 1D (TF1D) prediction of $\nu_z/\sqrt{2}$~\cite{Bus00}
where $\nu_z$ is the longitudinal trapping frequency. Our
experimental results are in quantitative agreement with numerical
simulations of the Gross-Pitaevskii equation (GPE).
They reveal that dark solitons can behave very similar to particles.
This is confirmed by explaining the essential features of the dynamics within a simple
physical picture regarding the dark solitons as particles in an
effective potential due to the external trap and their mutually
repulsive interactions. Being in the crossover regime, the role of
the transverse degrees of freedom has to be included in the
effective potential in order to get quantitative agreement between
theory and experiment~\cite{Geo07}.

Before elaborating on the theoretical models and systematic studies
we will briefly describe the details of the experimental setup.
We prepare a BEC of $^{87}$Rb in the $|F\!=\!2, m_F\!=\!2\rangle$ state containing about $N=1500$ atoms in a double well potential \cite{Alb05}. This potential is realized by  superimposing a far detuned crossed optical dipole trap ($\lambda = 1064$~nm)
and a one dimensional optical lattice ($\lambda = 843$~nm). The first beam
of the dipole trap has a gaussian waist of $5$~$\mu$m and results in
a strong transversal and weak longitudinal confinement.
The second beam orthogonally crosses the first one and has an elliptic shape ($60$~$\mu$m $\times$ $230$~$\mu$m waist) leading to an extra adjustable confinement only in the longitudinal direction of the trap. We start our experiments with a transversal frequency of the total harmonic trap of $\nu_{\perp} = 408$~Hz and a longitudinal one of $\nu_{z} = 63$~Hz. This corresponds to a power of about 1~mW in the first beam and 400~mW
in the second beam. The barrier height of the optical lattice is
chosen to be approximately 1~kHz and the lattice spacing is
$5.7$~$\mu$m. This results in a double well potential with a well
distance of $5.4$~$\mu$m.

In order to start with a well defined phase between the two
condensates the barrier height is chosen to be low enough such that
thermal phase fluctuations are negligible for the measured
temperature of $T \approx  10$~nK \cite{Rudi} (the critical
temperature for condensation is $T_c  \approx 110$~nK) and high
enough so that high contrast solitons are formed. Switching off the
optical lattice transforms the double well potential into a harmonic
trapping potential. Right after the switching off, we ramp to the
trap parameters of interest ($\nu_z, \nu_\perp$) and use an
optimized ramping time to minimize the excitation of quadrupole
oscillations (e.g. from ($\nu_z,\nu_{\perp}$)=(63~Hz,~408~Hz) to
(53~Hz,~890~Hz) within 10~ms for $N=1700$ atoms, or to (58~Hz,~408~Hz) within 3~ms for $N=950$). The distance
between the formed solitons can be adjusted by choosing different
sets of final trap frequencies. Imaging the atomic density after a
certain time of evolution in the harmonic trap is done using
standard absorption imaging. Although our optical resolution of
$\approx 1$~$\mu$m allows the direct observation of the solitons in
the trap, we use a short time of flight on the order of 1~ms to
enhance the contrast.

In our experiment the distance $D=5.4$~$\mu$m between the two
colliding condensates is well within the regime where the formation
of dark solitons is expected due to nonlinear interference. Dark
solitons are generated if the distance $D$ is smaller  than the
critical distance $D_c=\pi(6\frac{N\hbar a_s}{\nu_z m})^{1/3}=
25.8$~$\mu$m with $N$ being the number of atoms, $a_s$ the s-wave
scattering length, $\nu_z$ the longitudinal trap frequency and $m$
the atomic mass; if $D<D_c$, the interaction-energy exceeds the
kinetic energy and hence nonlinear dynamical phenomena are expected,
such as the formation of dark soliton pairs~\cite{Sco98}. This is
confirmed by 3D GPE simulations of the soliton formation and their
subsequent evolution in the trap as shown in Fig.~\ref{fig1}.
Including the optical and time resolution, the experimentally
observed density profile evolution is well reproduced. A dominant
pair of solitons oscillates close to the center of the cloud and we
can also distinguish additional pairs of solitons with much lower
contrast. In the following, we focus on the dynamics of the dominant
central pair and show that its oscillation frequency is well
described within a two soliton approximation.

\begin{figure}[h]
    \includegraphics[width=0.48\textwidth,keepaspectratio]{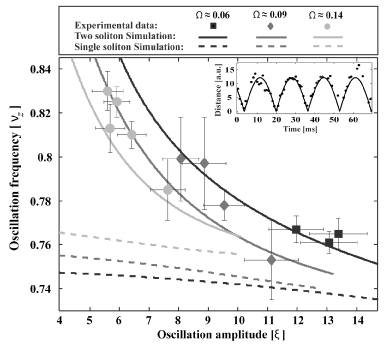}

    \caption{\label{fig3}
Comparison between experimentally obtained soliton oscillation
frequencies and NPSE simulation for one and two solitons. Each
frequency point is deduced from the temporal evolution of the
soliton distance as shown in the inset. Different symbols correspond
to different aspect ratios $\Omega$ of the trap. NPSE simulations
are represented by solid lines for the two soliton case, and by
dashed lines for the respective single soliton oscillations. The
error bars on the measured frequencies account for statistical
errors on the measured soliton and trap frequencies and systematic
errors on the atom number used to calculate the healing length.}
\end{figure}

We experimentally investigate the oscillation frequency of the
dominant soliton pair for different trap parameters and different
inter-soliton distances. A typical data set consists of 50 time
steps and 10 pictures per time step.
Our data does not allow to distinguish if the two solitons do or
do not cross at the collision point. Hence, for each time step we
measure the soliton distance which is well defined and reconstruct
its time evolution from which we extract an oscillation frequency as
shown in the inset of Fig.~\ref{fig3}. The obtained frequency is
divided by two in order to compare it to the oscillation frequency
expected for a single trapped soliton. The shot to shot
reproducibility of the soliton dynamics up to 100~ms allows the
observation of up to $7$ oscillation periods and hence the deduction
of the frequency with high accuracy. The typical statistical
experimental error in the frequency measurement is $\pm$1.5\%.
Fig.~\ref{fig3} shows the results of our frequency measurements and
their comparison with numerical simulations for the motion of two
trapped solitons using the Nonpolynomial Schr\"{o}dinger equation
(NPSE)~\cite{Sal02}. Note that this equation has been shown to be an
excellent approximation to the 3D GPE in the dimensionality
crossover regime where our experiments are performed, especially for
studying dark soliton dynamics~\cite{Geo07}.

In order to capture the essentials of the dynamics of the
experimentally realized soliton pairs, we initialize the condensate
with two solitons in the simulations such that the rms amplitude of
their oscillating motion matches the one observed experimentally.
The good agreement between numerics and experiments shows that the
dynamics produced by our method is well described within a two
soliton approximation even though extra solitons are produced. From
our experiment and the NPSE simulations, we observe an upshift up to
$16\%$ from the $\nu_z/\sqrt{2}$ prediction which was the first
value theoretically derived for the oscillation frequency of a
single trapped soliton~\cite{Bus00}. It is expected to be valid in a
1D trap in the asymptotic Thomas-Fermi limit ($N\Omega a_s/a_{\perp}
\ll 1$ and $((N / \sqrt{\Omega}) a_s/a_{\perp})^{1/3} \gg 1$)~\cite{Men02}
where $\Omega = \nu_z/\nu_{\perp} \ll 1$ is the aspect ratio of the
trap and $a_{\perp}$ the transverse harmonic oscillator length.

\begin{figure}[h]
    \includegraphics[width=0.48\textwidth,keepaspectratio]{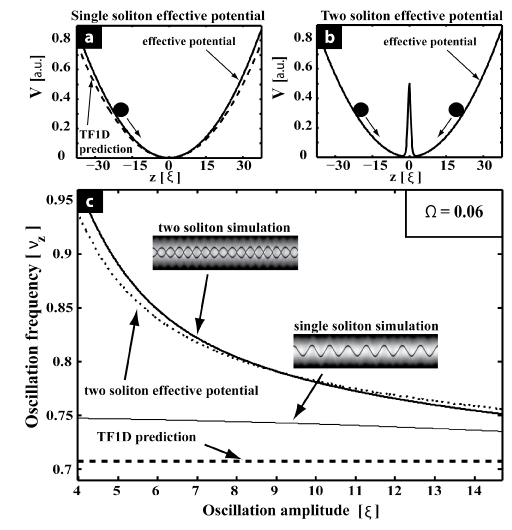}

    \caption{\label{fig2}
The oscillation dynamics of dark solitons in a trapped BEC is well
captured in an effective particle picture. For one soliton the
particle moves in a harmonic trap (a), while for two solitons an
additional barrier due to the repulsive interaction appears (b). The
dependence of the soliton oscillation frequency on the oscillation
amplitude from the trap center is shown in (c). The dashed line
shows the TF1D GPE result, the thin solid line indicates the upshift
mainly due to the dimensionality, while the thick solid line
includes the upshift due to the inter-soliton interaction obtained
by solving the NPSE. The dotted line represents the result obtained
by the simple effective particle model from Eqn.~(\ref{eqn1}).}
\end{figure}

We now give a theoretical description of the different effects
leading to the observed upshift for our situation of two oscillating
and interacting dark solitons including the dimensionality of the
trap. We consider the two solitons as particles moving in an
effective potential which arises from the combination of a harmonic
potential due to the trap \cite{Bus00} (see Fig.~\ref{fig2}a) and a
repulsive potential due to the interaction between the
solitons~\cite{Kiv95}. Because of the spatially symmetric
preparation, the effective potential is a symmetric double well
potential which is depicted in Fig.~\ref{fig2}b. This potential can
be expressed as a function of the distance $z$ of each of the
solitons from the trap center and its time derivative $\dot{z}$:
\begin{eqnarray}
V(z, \dot{z}) =  (2 \pi \nu_{1s})^2\, \frac{z^2}{2} + \frac{\mu
B^2}{2m \cosh^2(2 B z/\xi )}\label{eqn1}
\end{eqnarray}
where $B = \sqrt{1 - (\dot{z}/\xi)^2 (\hbar/\mu)^2}$ denotes the
darkness of the solitons, $\mu$ is a typical interaction energy on
the order of the chemical potential, $\xi= \sqrt{\hbar/(m \mu)}$ the
associated healing length and $\nu_{1s}$ the oscillation frequency
of a single trapped soliton. The frequency of the motion is obtained
by solving the Euler-Lagrange equation of motion associated with the
Lagrangian $ \mathcal{L}(z,\dot{z}) = \dot{z}^2/2 - V(z, \dot{z})$.
To obtain quantitative agreement, the model has to take into account
correctly both the free propagation of the solitons in the trap when
they are far away from each other ($z\gg\xi$) and the repulsive
interaction when they approach each other.

Good estimates for the single soliton parameter $\nu_{1s}$ and the
soliton interaction strength $\mu$ can be calculated as follows. The
single soliton frequency $\nu_{1s}$ is obtained by numerical
integration of the NPSE describing a single soliton. Because our
experimental parameters $\Omega  \approx 0.06 - 0.14$ and  $N \Omega
a_s/a_{\perp} \approx 1.2 - 1.8$ are both in the crossover regime
and far from the Thomas-Fermi limit, substantial corrections to the
asymptotic value $\nu_{z}/\sqrt{2}$ are expected. Therefore the
oscillation frequency of a single dark soliton is upshifted by a few
percent from the asymptotic value as discussed in detail using the
Bogoliubov-de Gennes analysis of the NPSE in~\cite{Geo07} (see
Fig.~\ref{fig2}a).
Our simulations reveal that for the example of
our parameter sets with $\Omega  \approx 0.06$ this upshift is 5\%
(see Fig.~\ref{fig2}c). Note that the frequencies obtained by 1D GPE
simulations are approximately 2\% higher than the asymptotic limit
because the Thomas-Fermi limit is not reached for our experimental
parameters as discussed in \cite{Kon03}.
The effect of dimensionality of the system, i.e. the
role of the transverse degrees of freedom which is captured only by
the NPSE or the 3D GPE, accounts for the remaining 3\%.
Fig.~\ref{fig2}c shows the comparison between the asymptotic limit
$\nu_{z}/\sqrt{2}$ and the single soliton NPSE simulation for one
specific trap. The simulation results for the three different parameter
sets used in the experiment are shown in Fig.~\ref{fig3}. As
expected, the single soliton frequency increases with the aspect
ratio.

As shown in Fig.~\ref{fig2}c, the repulsive interaction between the
solitons results in an upshift of the oscillation frequency,
compared to the single soliton case, that strongly depends on the
oscillation amplitude. Our model accurately reproduces the upshift
if the interaction parameter $\mu$ is set to be the chemical
potential of the condensate obtained from the 3D GPE equation. In
our experimentally accessible parameter range, the agreement of the
model with NPSE simulations is better than  5\%. This allows us to
clearly identify the significant role of the repulsive interactions
and shows that the effective repulsive potential in
Eqn.~(\ref{eqn1}) obtained in the 1D homogeneous case is a good
approximation to our complex situation.

In conclusion we controllably create pairs of dark solitons by
colliding two atomic clouds released from a double well potential in
a harmonic trap. The full dynamics of multiple dark soliton oscillations
and collisions can be observed for the first time, allowing for precise
frequency measurements and showing that dark solitons
are still stable after several collisions.
The experimentally observed total upshifts from the TF1D frequency prediction are up to 16\%. A simple effective particle picture confirms that the final oscillation
frequency of two solitons is affected by two effects namely the
single soliton frequency upshift and the inter-soliton interaction.
The presented robust method for preparing solitonic excitations will
be a starting point for further studies of dark soliton dynamics in
the presence of designed potentials as well as for a possible route
towards multi-soliton interaction and perhaps even dark soliton gases.

\begin{acknowledgments}
We would like to thank Peter Schmelcher for discussions at the early
stages of this project, and B\"orge Hemmerling, Rudolf Gati and Timo
Ottenstein for their help in the experimental setup. We gratefully
acknowledge support from NSF, DFG and from the Alexander von
Humboldt Foundation. J.E. aknowledges support from the EC
Marie-Curie program (EIF-040721).

\end{acknowledgments}


\bibliography{DS_bibfile_080319short}

\end{document}